\documentclass[onecolumn]{aa}  
\usepackage{verbatim}
\usepackage{color}
\newcommand       \K             {\,{\rm K}}

\newcommand       \mum      {\,{\rm \mu m}}

\newcommand       \mus       {\,{\rm\mu s}}

\newcommand       \eV          {\,{\rm eV}}
\newcommand       \keV          {\,{\rm keV}}

\newcommand     \gtsim  {\lower.5ex\hbox{$\buildrel > \over \sim$}}
\newcommand     \ltsim  {\lower.5ex\hbox{$\buildrel < \over \sim$}}
\newcommand     \simgt  {\lower.5ex\hbox{$\buildrel > \over \sim$}}
\newcommand     \simlt  {\lower.5ex\hbox{$\buildrel < \over \sim$}}
\newcommand       \simali       {{\sim\,}}
\begin{document}

\titlerunning{Fragmentation and isomerization of coronene}
\title{Fragmentation and isomerization of
         polycyclic aromatic hydrocarbons
         in the interstellar medium:
         coronene as a case study}

\authorrunning{Chen, Luo \& Li}
   \author{Tao Chen
          \inst{1}\fnmsep\thanks{Email: taochen@kth.se}
          \and
          Yi Luo
          \inst{1,2}\fnmsep\thanks{Email: luo@kth.se}
          \and
          Aigen Li
          \inst{3}\fnmsep\thanks{Email: lia@missouri.edu}
          }
          
   \institute{
         School of Engineering Sciences in Chemistry, Biotechnology and Health, Department of Theoretical Chemistry and Biology, Royal Institute of Technology, 10691, Stockholm, Sweden
         \and
         Hefei National Laboratory for Physical Science at the Microscale, Department of Chemical Physics, School of Chemistry and Materials Science, University of Science and Technology of China, Hefei, 230026 Anhui,  China
         \and
         Department of Physics and Astronomy, University of Missouri, Columbia, MO 65211, USA
          }

\abstract
  {}
     {Due to the limitations of current computational technology, the fragmentation and isomerization products of vibrationally-excited polycyclic aromatic hydrocarbon (PAH) molecules and their derivatives are poorly studied. In this work, we investigate the intermediate products of PAHs and their derivatives as well as the gas-phase reactions relevant to the interstellar medium, with coronene as a case study.}
     {Based on the semi-empirical method of PM3 as implemented in the CP2K program, molecular dynamics simulations are performed to model the major processes (e.g., vibrations, fragmentations, and isomerizations) of coronene and its derivatives (e.g., methylated coronene, hydrogenated coronene, dehydrogenated coronene, nitrogen-substituted coronene, and oxygen-substituted coronene) at temperatures of 3000$\K$ and 4000$\K$.} 
     {We find that the anharmonic effects are crucial for the simulation of vibrational excitation. For the molecules studied here, H$_2$, CO, HCN, and CH$_2$ are the major fragments. Following the dissociation of these small units, most of the molecules could maintain their ring structures, but a few molecules would break completely into carbon chains. The transformation from hexagon to pentagon or heptagon may occur and the heteroatomic substitutions (e.g., N- or O-substitutions) facilitate the transformation.}
     {}

   \keywords{astrochemistry, molecular data, molecular processes, ISM: molecules, methods: laboratory: molecular, (ISM:) photon-dominated region (PDR)} 

   \maketitle

\section{Introduction}
Polycyclic aromatic hydrocarbon (PAH) molecules are commonly thought to be responsible for the distinctive set of aromatic infrared (IR) emission features at 3.3, 6.2, 7.7, 8.6 and 11.3$\mum$ seen in a wide variety of astrophysical environments \citep{leger1984identification, allamandola1985polycyclic}. In the interstellar medium (ISM), PAHs are vibrationally excited by ultraviolet (UV) and visible stellar photons \citep{li2002infrared}. Following the photo-excitation, electronic relaxation takes place predominantly by nonradiative processes, in which the intramolecular vibrational energies are redistributed randomly over all accessible vibrational degrees of freedom of the electronic ground state of the molecules \citep{allamandola1989interstellar}. The vibrational de-excitation occurs from these vibrationally hot molecules, which lead to the aromatic IR emission bands \citep{tielens2008}. Besides the IR emission, fragmentation is a competing process for a vibrationally excited molecule. However the timescales are very different for these two processes. The fragmentation resulted from photo-driven processes takes place in a timescale of picoseconds (ps), and the loss of H, H$_2$, or C$_{2n}$H$_{x}$ usually dominates the experimental mass spectra of PAH fragmentation \citep{west2014photodissociation, chen2018planes, zhen2018laboratory}. As the redistribution of the absorbed energy and the subsequent fragmentations are random, such a process is usually called statistical fragmentation \citep{chen2014absolute, chen2015}. 

In addition to photo-driven processes, PAHs are also subjected to energetic ions from stellar winds or supernova explosions \citep{micelotta2010shocks}. Motivated by this astrophysical interest, the ion-induced fragmentation and molecular growth processes have recently been extensively investigated \citep{zettergren2013formations, seitz2013ions, stockett2014fragmentation, stockett2014nonstatistical, chen2014absolute, delaunay2015molecular, chen2015}. At low energies (with center-of-mass energy less than 1\,keV), energetic ions lead to the loss of H and C atoms. The chemical routes and fragmentation products are expected to appreciably differ from that of photo-driven processes (i.e., the statistical fragmentation). This process is called non-statistical fragmentation \citep{stockett2014nonstatistical, chen2014absolute, chen2015}, in which, single C-losses are commonly observed \citep{stockett2014nonstatistical, stockett2014fragmentation, stockett2015isomer}. Non-statistical fragmentation occurs approximately thousand times faster than the statistical fragmentation, i.e., at a timescale of femtoseconds (fs).  

At high center-of-mass energies ($\simgt10\keV$) of ion-induced dissociation, the statistical fragmentation is dominant, i.e., H-, H$_2$-, and C$_{2n}$H$_{x}$-losses are most commonly detected on the experimental mass spectra \citep{zettergren2013formations, seitz2013ions, delaunay2015molecular, chen2015}. This process is crucial for a PAH molecule to reach high internal energy or temperature, which can not be reached through the absorption of single photons below the Lyman limit. The main drawback with the experiments is that only the mass-to-charge ratios can be measured and no detailed information about the products, the intermediate fragments, and the reaction processes can be recorded. To compensate such a drawback, static quantum chemical calculations, e.g., density functional theory (DFT), are performed to reveal the possible structures and reactions associated with a product. The reaction pathways for the formation of H and H$_2$ from PAHs have been revealed through static DFT \citep{paris2014multiple, chen2015, castellanos2018photoinduced}. However, the static methods rely heavily on the empirical knowledge about complex systems/reactions, e.g., the mechanism for C$_{2n}$H$_{x}$-losses or complete fragmentation of carbon skeletons, can hardly be explored using the static methods, instead, molecular dynamics simulations are widely performed to investigate the detailed fragmentation processes \citep{martin2016post, krasnokutski2017growth,trinquier2017pah,simon2018dissociation,rapacioli2018atomic,chen2019formation}.  

In this work, we apply molecular dynamics simulations 
to investigate the vibrations, H-migrations, dissociations, 
and isomerizations of PAHs, 
with coronene {(C$_{24}$H$_{12}$),   
a highly symmetric PAH species,
as a case study.\footnote{%
  The 3.3$\mum$ PAH emission band,
   commonly seen even in UV-intense reflection nebulae, 
  planetary nebulae, and H{\sc ii} regions \citep{tielens2008}, 
  arises predominantly from small PAHs of 
  $\simali$20--30 carbon atoms 
  [see Figure~7 of \cite{draine2007infrared}]. 
  Exposed to the intense UV radiation 
  in these regions, compact, 
  pericondensed PAH molecules 
  like coronene, ovalene, 
  circumcoronene are more likely 
  to survive than catacondensed PAHs 
  with an open, irregular structure. 
  Therefore, in this work we consider 
  coronene as a case study.
  }

In astronomical environments, 
PAH molecules may include substituents 
like N in place of C
(e.g., see \citealt{hudgins2005variations}, \citealt{mattioda2008near})
and O in place of C 
(e.g., see \citealt{bauschlicher1998infrared}).
Observationally,  the C--C stretch 
peaks at wavelengths as short as 6.2$\mum$, 
in contrast,
both experimental and computational spectra
of {\it pure} PAHs reveal that it often occurs 
at $\simgt$\,6.3$\mum$.
The subtle variations in the peak wavelength of 
the 6.2$\mum$ C--C emission band
are  commonly attributed to
polycyclic aromatic nitrogen heterocycles
--- PAHs with one or more nitrogen atoms 
substituted into their carbon skeleton \citep{hudgins2005variations, mattioda2008near}.
In regions with intense UV radiation, 
small PAHs with fewer than 25 carbon atoms
are expected to be partially dehydrogenated 
(e.g., see \citealt{tielens1987shock}, 
\citealt{malloci2008dehydrogenated}).\footnote{%
   More recently, \cite{andrews2016hydrogenation} modeled
   the physical and chemical processes of coronene,
   circumcoronene and circumcircumcoronene in
   the north-west photodissociation region (PDR) 
   of the reflection nebula NGC~7023
   where the UV starlight intensity is $\simali$2600
   times that of the local ISM. They found
   that coronene would be fully dehydrogenated
   in the NGC~7023 PDR.
    \cite{montillaud2013evolution} found 
    that coronene would be fully dehydrogenated even in the diffuse ISM.   
    }
In H-rich, UV-poor benign regions,
PAHs may be superhydrogenated
and their edges contain excess H atoms
(e.g., see \citealt{bernstein1996hydrogenated}, \citealt{thrower2012experimental}, \citealt{sandford2013infrared}).
They may also contain one or several 
methyl sidegroups as revealed by
the detection in many PAH sources
of a weak satellite emission feature
at 3.4$\mum$ which always accompanies
the 3.3$\mum$ emission feature
(see \citealt{yang2017carriers} and references therein).
Finally, experiments have shown that 
UV photolysis of low-temperature 
coronene-ice mixtures in dense molecular clouds
would lead to the addition to PAHs
of various functional groups 
\citep{bernstein1999uv, gibb2000inventory, bernstein2002}, 
including methyl (--CH$_3$), 
methoxy (--OCH$_3$), 
cyano (--CN),
isocyano (--NC), 
alcohol (--OH), and 
ketone ($>$C=O).

To explore the influence of the structural
complexities of PAHs on their fragmentation 
and isomerization processes, 
we therefore also examine 
seven derivatives of coronene,
including methylated coronene,
hydrogenated coronene, 
dehydrogenated coronene, 
N-substituted coronene,
O-substituted coronene,
carbonylcoronene
and methoxycoronene 
(see Figure~\ref{fig:m8}).

This paper is organized as follows. 
Section~2 shows the importance of the anharmonic effects, 
without such effects taken into account neither dissociation 
nor isomerization could be accurately studied. 
Section~3 elaborates the various relevant 
fragmentation processes. 
In Section~4 we show a new H roaming route. 
Section~5 explores the formation of pentagon 
and heptagon in PAH derivatives. 
Finally, we summarize the major results in Section~6. 

%%% Figure 1 %%%
\begin{figure}
\begin{center}
\includegraphics[width=1.0\columnwidth]{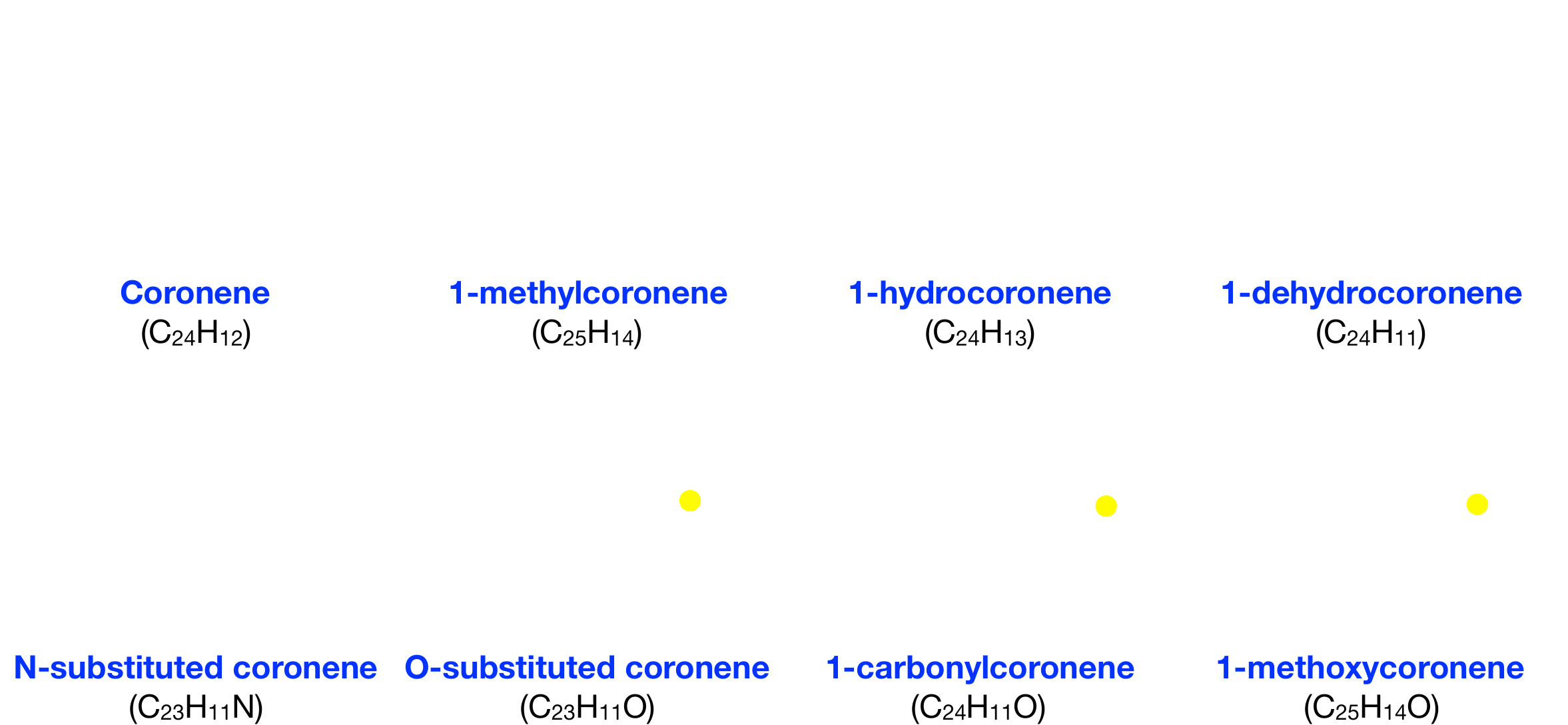}
\caption{Structures of coronene and its seven derivatives studied in this work. The names and chemical formulae are shown beneath each molecule.}
\label{fig:m8}
\end{center}
\end{figure}

\section{Anharmonicity}
For harmonic oscillators, the vibrational energy levels on the parabolic potential energy surface are equally separated (see Figure~\ref{fig:pes} for illustration), which can be calculated from the following: 
\begin{equation}
E_{\rm har}(n) = \sum_{i=1}^{3N-6} h\nu_i \left(n_i + \frac{1}{2}\right)~~, 
\label{eq1}
\end{equation}
where $h$ is the Planck constant, $\nu_i$ is the frequency of the $i$-th vibrational mode (for a non-linear molecule of $N$ atoms, there will be $3N-6$ vibrational modes),  and $n_i \equiv$ ($n_1$, $n_2$, ...) is the quantum number representing the state of each vibrational mode. The fundamental vibrational frequency of a molecule corresponds to the transition from the ground level ($n = 0$) to the first vibrational excitation level ($n = 1$). The zero point energy is $E_{\rm har}(0) = \left(1/2\right) \sum_i h \nu_i$. The transitions from $n=0$ to $m>1$ are called overtones, and the transitions from $n \ge 1$ to  $m > n$ are called hot transitions or hot bands. 

According to eq.\,\ref{eq1}, a molecule with harmonic bonds could reach infinite energy $E_{\rm har}(n \rightarrow \infty$) without bond breaking. However, this is not the case in reality. Molecules do dissociate at highly excited vibrational states \citep{chen2015, chen2019formation}. The potential energy surface is not a perfectly parabolic shape, instead, it is a non-symmetric open well (i.e., anharmonic potential, see Figure~\ref{fig:pes}), which allows a molecule to break up above certain vibrational states. On the anharmonic potential energy surface, the energy levels are unequally separated (see Figure~\ref{fig:pes} for illustration), which can be estimated from the following formula \citep{burcl2003vibrational}:
\begin{equation}
E_{\rm anh}(n) = \sum_i h\nu_i \left(n_i + \frac{1}{2}\right) 
+ \sum_{i \le j}\chi_{ij}\left(n_i + \frac{1}{2}\right) 
\left(n_j + \frac{1}{2}\right) ~~,
\label{eq2}
\end{equation}
where $\chi_{ij}$, the anharmonic coupling which describes the interactions (mode couplings) or resonances among various vibrational modes, can be approximated as a 2-dimensional matrix, and most of its elements are negative, which leads to the anharmonic energy levels $E_{\rm anh}(n)$ lower than the corresponding harmonic energy levels $E_{\rm har}(n)$. Moreover, due to mode couplings, combination bands (two or more fundamental vibrations are excited simultaneously) show up, and the positions and intensities of the fundamental bands also change \citep{chen2018carrier}. The band shifting and broadening have been seen in the vibrational spectra calculated by molecular dynamics simulations \citep{chen2019temperature}, suggesting that anharmonicity and mode couplings are intrinsically included in the molecular dynamics simulations.

%%% Figure 2 %%%
\begin{figure}
\begin{center}
\includegraphics[width=1.0\columnwidth]{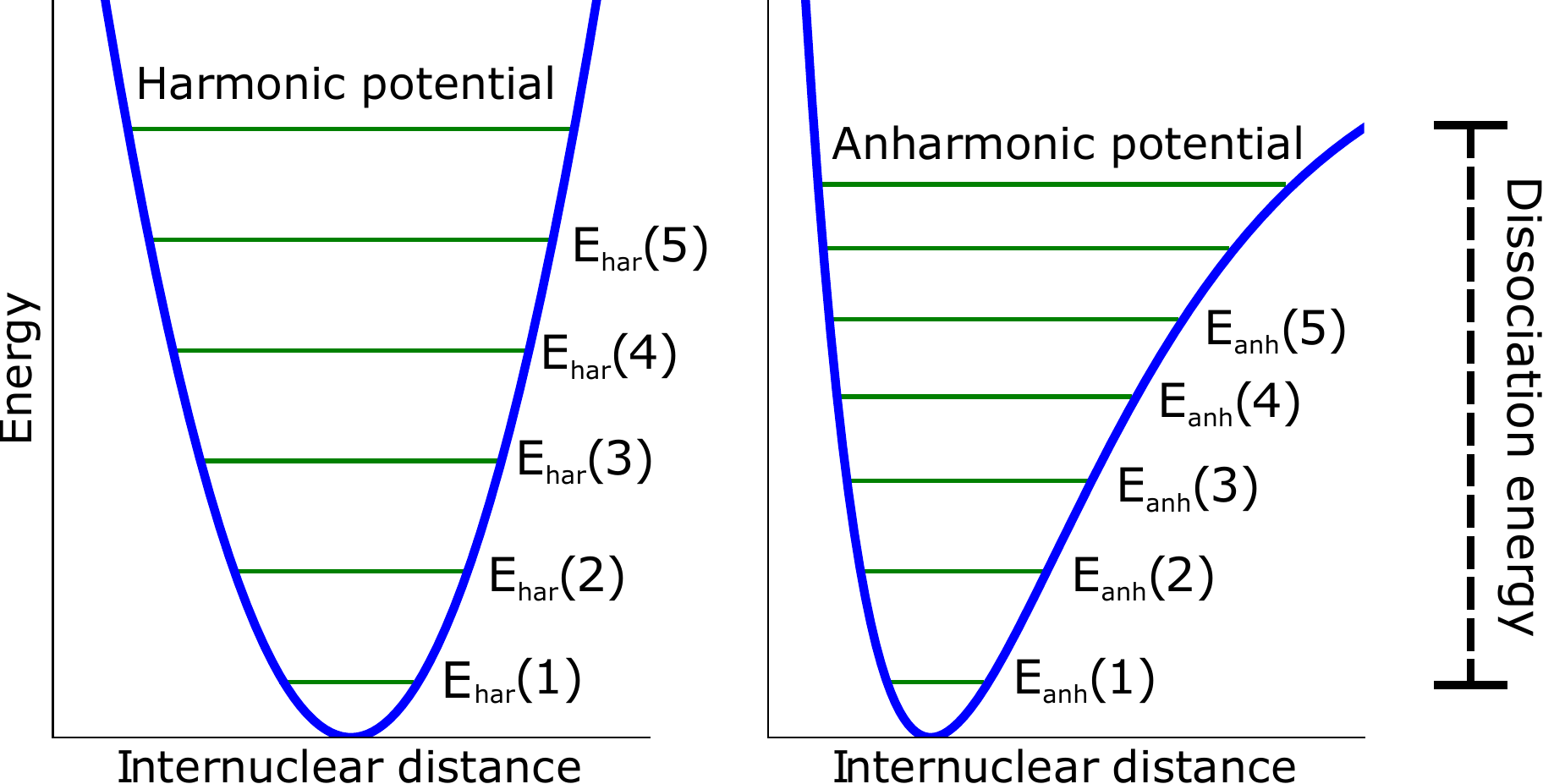}
\caption{Harmonic and anharmonic potentials. Harmonic potentials can be described by a parabolic function, which is a symmetric well and closes in both sides. A molecule does not break up with such a harmonic potential. Anharmonic potential is an asymmetric and open well. At certain vibrational states, a molecule dissociates.}
\label{fig:pes}
\end{center}
\end{figure} 

\section{Fragmentation}
Figure~\ref{fig:expcoronene} shows the experimental mass spectrum of coronene following collisions with He$^+$ \citep{chen2015}. Large amounts of H losses and fragmentations of the carbon skeleton are clearly seen. The mechanism for H losses has been studied in detail \citep{chen2015, castellanos2018photoinduced}. However, due to the complex isomerization and dissociation features at highly excited vibrational states, the reaction pathway for the fragmentation of the carbon skeleton remains uncharted. Here, we investigate such reactions using molecular dynamics simulations. 

%%% Figure 3 %%%
\begin{figure}
\begin{center}
\includegraphics[width=1.0\columnwidth]{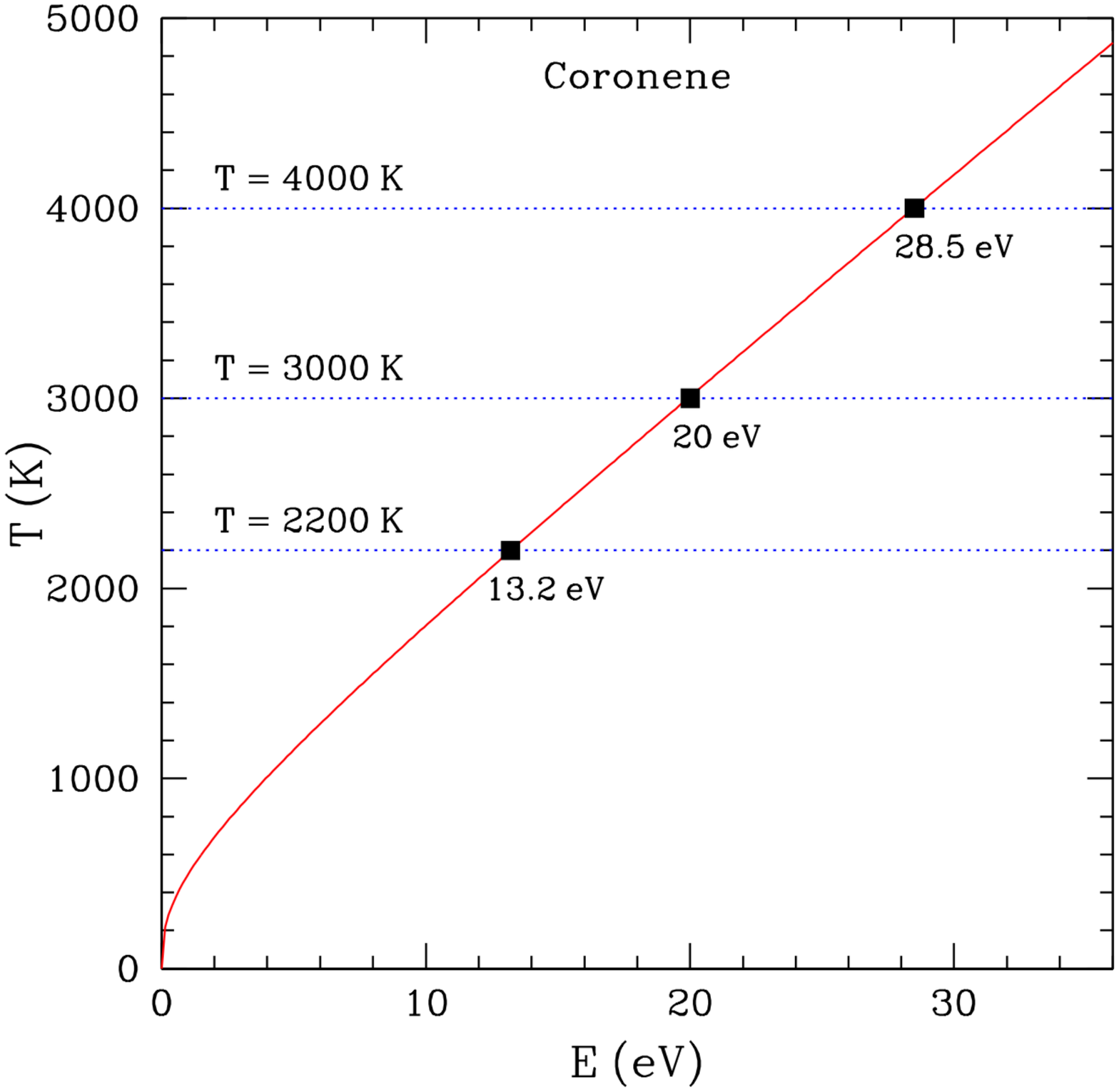}
\caption{Temperature $T$ for coronene with vibrational energy $E$, calculated from its vibrational mode spectrum (see Draine \& Li 2001). For coronene to attain a vibrational temperature of $T$\,$\simgt$\,2200, 3000 and 4000$\K$, its vibrational energy $E$ has to exceed, respectively, $\simali$13.2, 20.0, and 28.5$\eV$. This can only be achieved for PAHs excited by multiple photons or energetic particles.}
\label{fig:TvsE}
\end{center}
\end{figure}

It has been reported that the fragmentation only becomes important for temperatures exceeding $\simali$2200$\K$, regardless of the PAH size and the excitation agent \citep{chen2015}. Therefore, we simulate vibrational excitations starting from 3000$\K$. We should note that, as shown in Figure~\ref{fig:TvsE}, for a coronene molecule at a vibrational temperature of $T$\,$\simgt$\,3000$\K$, its vibrational energy has to exceed $\simali$20$\eV$. This indicates that the fragmentation process described here would be mostly applicable to UV-intense regions where PAHs are excited by multiple photons or hostile environments where PAHs are collisionally excited by energetic particles. As the energy can be randomly redistributed across all degrees of freedom, such simulations correspond to statistical fragmentations. The simulations are performed using the Quickstep module of the CP2K program package \citep{vandevondele2005quickstep}. The semi-empirical method of PM3 \citep{stewart1989optimization} is utilized for electronic structure calculations. It has been shown that such methods are capable of producing reasonable vibrational spectra for large compact PAHs \citep{chen2019temperature} and high-temperature dissociation pathways for linear PAHs \citep{chen2019formation}. Canonical ensembles (NVT) are performed at multiple temperatures with the Nos\'e-Hoover chain thermostat \citep{nose1984unified, nose1984molecular, martyna1992nose}. Starting from optimized geometries of the individual molecules, the systems are equilibrated for 10\,ps. The production simulations are subsequently run for 50\,ps, at a time step of 0.1\,fs. 

%%% Figure 4 %%%
\begin{figure}
\begin{center}
\includegraphics[width=1.0\columnwidth]{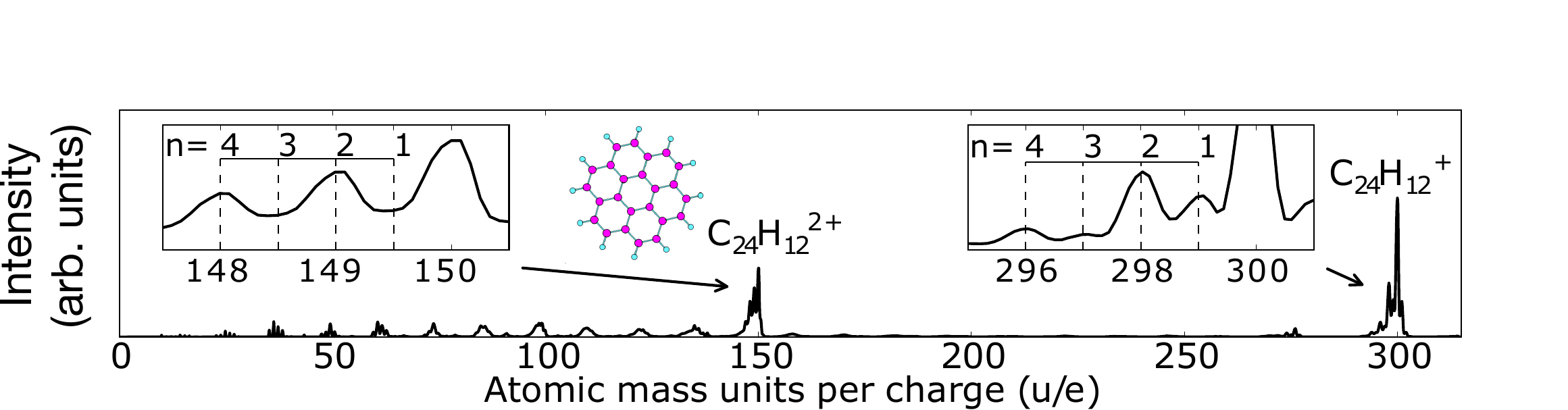}
\caption{Mass spectrum of coronene following collisions with 11.25$\keV$ He$^+$. The highest peaks correspond to the single and double ionizations of coronene without fragmentation on the experimental time scale of $\simali$10$\mus$. The left and right zoom-ins respectively show the intensity distributions for the losses of different numbers, $n$, of H atoms from the coronene dications and cations. The loss of an even number of H atoms is more favorable than the loss of an odd number of H atoms. Other peaks are due to the fragmentation of the carbon skeleton in one or several steps. See \cite{chen2015} for details.}
\label{fig:expcoronene}
\end{center}
\end{figure}

Figure~\ref{fig:fragmentation_3000K} shows the snapshots from molecular dynamics simulations for 1-hydrocoronene and O-substituted coronene at 3000$\K$. At such a temperature, no fragment is observed for other molecules. As demonstrated in Figure~\ref{fig:fragmentation_3000K}, due to the addition of an extra H atom to a peripheral carbon, the corresponding C--H bond converts from aromatic to aliphatic, which weakens the strength of the connection and eventually leads to the first break up of the C--H bond at 0.28\,ps. In comparison to a pristine coronene which breaks up the first C--H bond at 4000\,K, the hydrogenated coronene is less stable. This phenomenon is consistent with a recent finding that additional H atoms do not protect PAH molecules from fragmentation \citep{gatchell2015failure}. At 0.54\,ps a hydrogen atom migrates from the body of the newly formed carbon chain to the unsaturated carbon, and at 0.55\,ps a new C--H bond is formed. At 1.26\,ps, a hydrogen atom on the head of the carbon chain is dissociated, i.e., sp$^3$ C--C or sp$^2$ C=C bonds tend to form sp C$\equiv$C bonds by removing H atoms \citep{chen2019formation}. At the end of the simulation (50\,ps), a smaller PAH (with only one ring broken) attached with an --C$_2$H sidegroup is formed, which agrees with the results reported recently for linear PAHs \citep{chen2019formation}.

For the O-substituted coronene, one of the C--O bond breaks up at first at 8.08\,ps. Similar to 1-hydrocoronene, the hydrogen atom linked to the other side of the C--O bond migrates to the unsaturated carbon to form a new C--H bond. A CO molecule is formed at 13.82\,ps and dissociates from the maternal molecule. No other fragment is observed and no ring is broken until the end of the simulation (50\,ps). This process is consistent with the experimental and theoretical studies of the photodissociation of bisanthenequinone cations, in which the CO-loss was found to be the lowest dissociation channel \citep{chen2018planes}. 

%%% Figure 5 %%%
\begin{figure}
\begin{center}
\includegraphics[width=1.0\columnwidth]{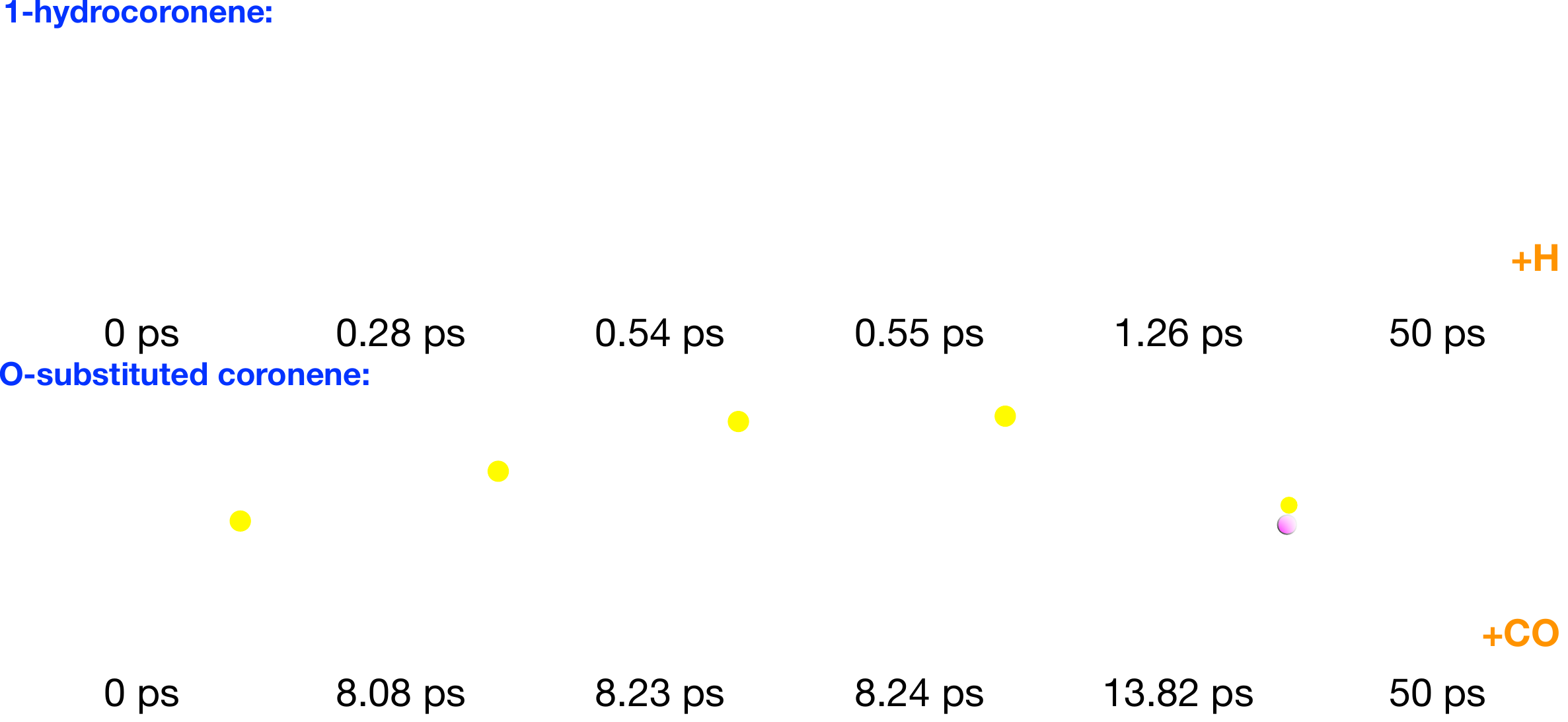}
\caption{Fragmentation of 1-hydrocoronene (top) and O-substituted coronene (bottom) at 3000\,K. The orange text in the cells indicates the fragment(s) which are located far away from the maternal molecule.}
\label{fig:fragmentation_3000K}
\end{center}
\end{figure}

The other molecules dissociate at 4000\,K. 
Again, as shown in Figure~\ref{fig:TvsE}, such a high vibrational temperature can only be achievable for 
PAHs vibrationally excited by multiple photons or energetic particles. 
Figure~\ref{fig:fragmentation_4000K} shows the fragmentation processes of coronene, 1-methylcoronene, coronene with a H-loss,
N-substituted coronene, coronene with a hydrogen atom
substituted by an oxygen atom, and 1-methoxycoronene. 
For coronene, a H$_2$-loss is observed, it is very similar to the reaction pathway as
reported in the literature, i.e., H$_2$ is formed via H migration
without breaking the ring \citep{paris2014multiple, chen2015,
  castellanos2018photoinduced}. For 1-methylcoronene, the C--CH$_3$
bond breaks at 0.25\,ps. In the meantime, one of the hydrogen atoms on
the methyl group moves to the carbon atom connected with --CH$_3$. At
0.31\,ps, a bare coronene molecule is formed and a CH$_2$-loss is
observed simultaneously. No more fragment is found until the end of
the simulation. However, one of the peripheral hexagons tranforms to a
pentagon attached with a --CH$_2$ sidegroup (see the last cell in the second row of Figure~\ref{fig:fragmentation_4000K} for details). 

Both coronene and 1-methylcoronene maintain the pristine aromatic structures, i.e., most of the rings do not break at 4000\,K. However, the situation is different for the coronene with a H-loss. At 1.46\,ps, three rings are broken to carbon chains, and a hydrogen atom is released. At 3.58\,ps, all the rings are broken, and a C$_5$H$_2$ molecule and a longer carbon chain are dissociated from each other. Subsequently, a H-loss from the longer chain is observed at 5.22\,ps. At 6.59\,ps, a C$_3$H$_2$ molecule is separated from the longer chain. The molecule is fragmented completely to multiple carbon chains and H atoms. 

For the N-substituted coronene, HCN is formed at 0.425\,ps. No more fragment is found until the end of the simulation. However, the maternal molecule isomerizes to a smaller PAH attached with carbon chains. 
The fragmentation of the coronene with a hydrogen atom substituted by an oxygen atom is also violent. At 0.585\,ps, five rings are broken, and a CO unit is released from the main body. At 0.815\,ps, a hydrogen atom is dissociated. At 1.275\,ps, a C$_5$H$_2$ molecule is separated. Similar to the case for the coronene with a H-loss, the molecule is dissociated completely to multiple carbon chains, CO and H atoms. 
One-methoxycoronene is the largest molecule in this study, which contains 40 atoms. At 6.66\,ps, a H$_2$ molecule is released from the methoxyl group. At 15.07\,ps, another H$_2$ molecule is dissociated and the carbon atom on the methoxyl group migrates from the oxygen atom to a neighboring carbon atom, i.e., a --CO unit is formed. As discussed above for both O-substituted coronene and 1-carbonylcoronene, the C--CO bond always breaks before C--H and C--C bonds. At 21.84\,ps, a CO-loss is observed. Thereafter, no more fragment is found and no rings break until the end of the simulation. 

%%% Figure 6 %%%
\begin{figure*}
\begin{center}
\includegraphics[width=0.9\textwidth]{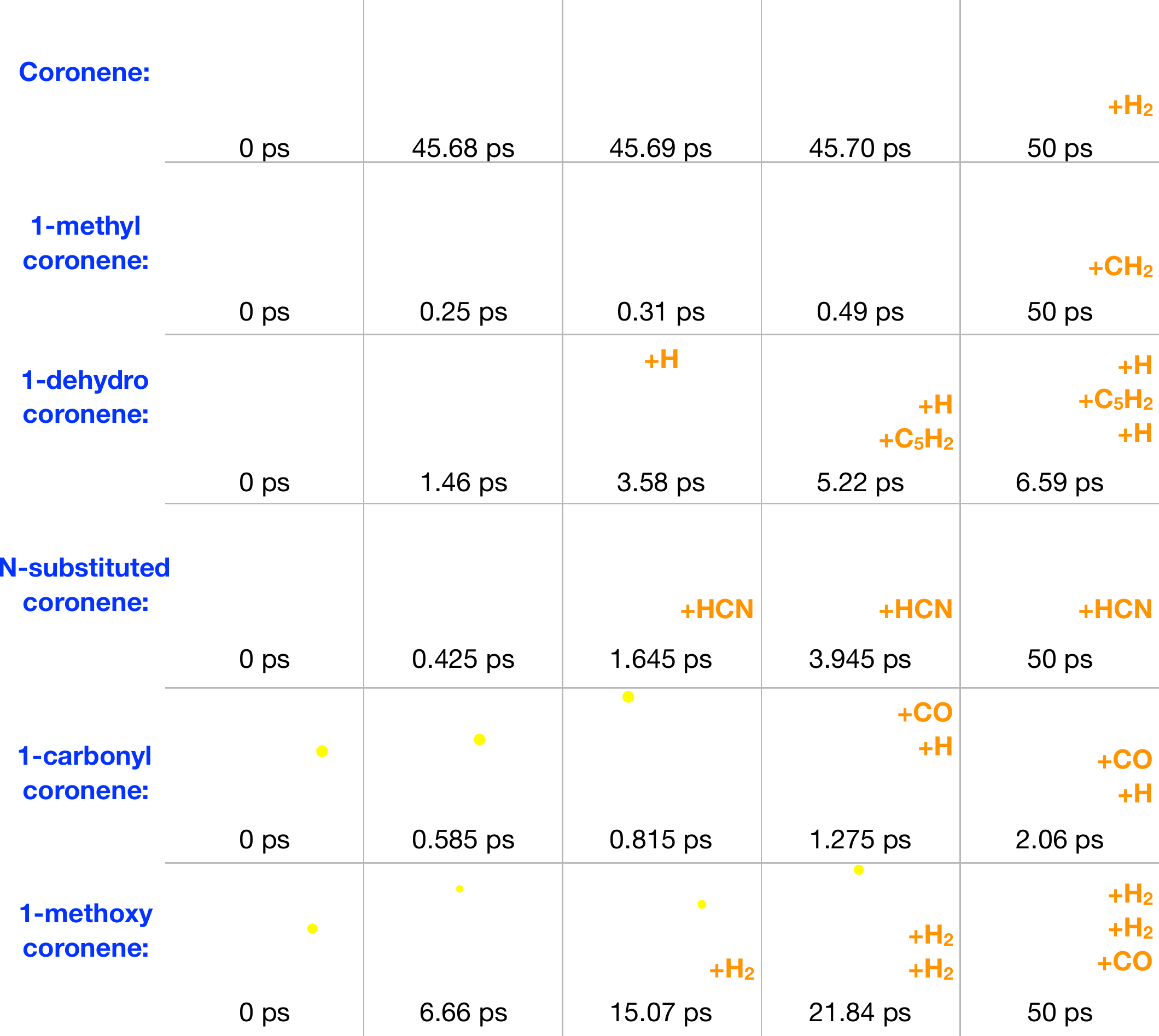}
\caption{Fragmentation of (from top to bottom) coronene, 1-methylcoronene, coronene with a H-loss, nitrogen-substituted coronene, 1-carbonylcoronene and 1-methoxycoronene. The simulations are performed at 4000\,K. The orange text(s) in the cells indicate the fragment(s) which are located far away from the maternal molecule.}
\label{fig:fragmentation_4000K}
\end{center}
\end{figure*}

\section{Hydrogen roaming}
The roaming of hydrogen atoms is crucial for the formation of aliphatic bonds and H$_2$ molecules \citep{paris2014multiple, chen2015, castellanos2018photoinduced}. Previous studies have reported hydrogen roaming on the periphery of PAHs. In this study, we find a new roaming route. Figure~\ref{fig:h-migration} shows the hydrogen roaming on the carbon skeleton of N-substituted coronene. Two hydrogen migrations are demonstrated: at 4.53\,ps, one hydrogen atom is located at the end of a carbon chain, the other one is bound to an inner carbon atom belonging to a pentagon. At 4.545\,ps, the carbon atom located at the end of the carbon chain dissociates from the carbon chain, and at 4.55\,ps, it migrates to a nearby unsaturated carbon atom belonging to a hexagon. At 4.62\,ps, the hydrogen atom located at the inner carbon atom migrates to the carbon chain which transfers a hydrogen atom to a nearby carbon atom at 4.55\,ps. At 4.635\,ps, all the peripheral carbon atoms become saturated with hydrogen atoms.

%%% Figure 7 %%%
\begin{figure}
\begin{center}
\includegraphics[width=1.0\columnwidth]{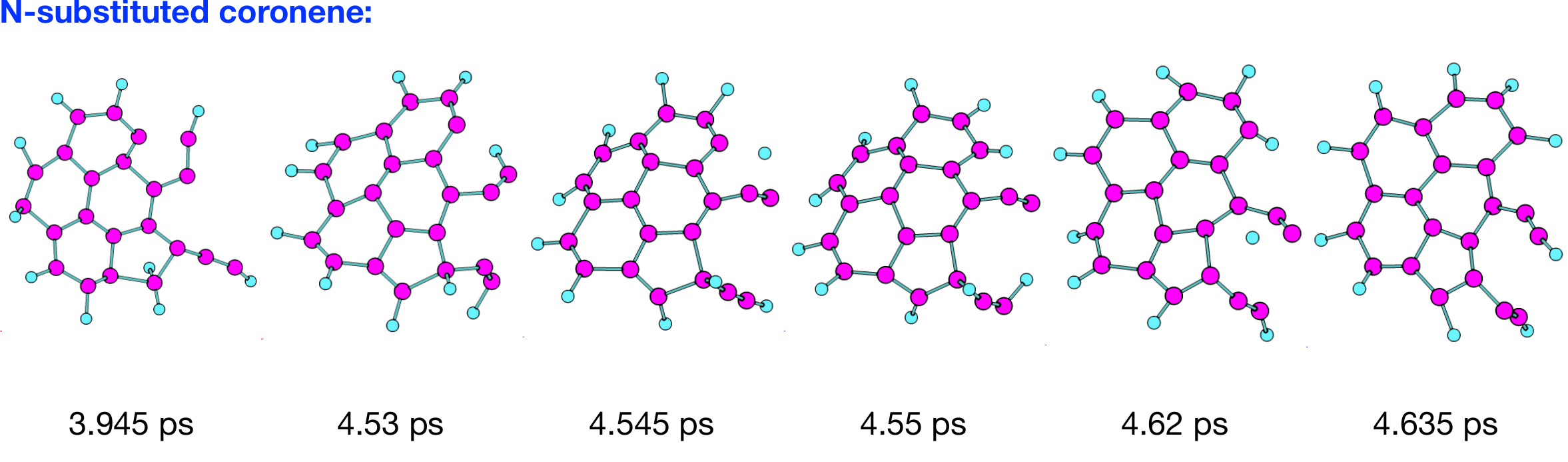}
\caption{Hydrogen roaming on C$_{22}$H$_{10}$ fragment produced by ejection of HCN from nitrogen-substituted coronene.}
\label{fig:h-migration}
\end{center}
\end{figure}  

\section{Formation of pentagon and heptagon}
Figure~\ref{fig:bq} shows the formation of pentagons in the photodissociation of bisanthenequinone cations (Bq$^+$), in which the Bq$^+$ cation does not dehydrogenate, but instead fragments via the sequential loss of two neutral carbonyl groups (--CO), causing the formation of pentagons. In addition, the molecule transforms into a bowl-shaped one following the loss of the second --CO group. Such process is proposed to be an important step for the formation of fullenerne in the ISM \citep{chen2018planes}. Similar reactions can also be found in Figure~\ref{fig:fragmentation_4000K} and Figure~\ref{fig:h-migration}. 

%%% Figure 8 %%%
\begin{figure}
\begin{center}
\includegraphics[width=1.0\columnwidth]{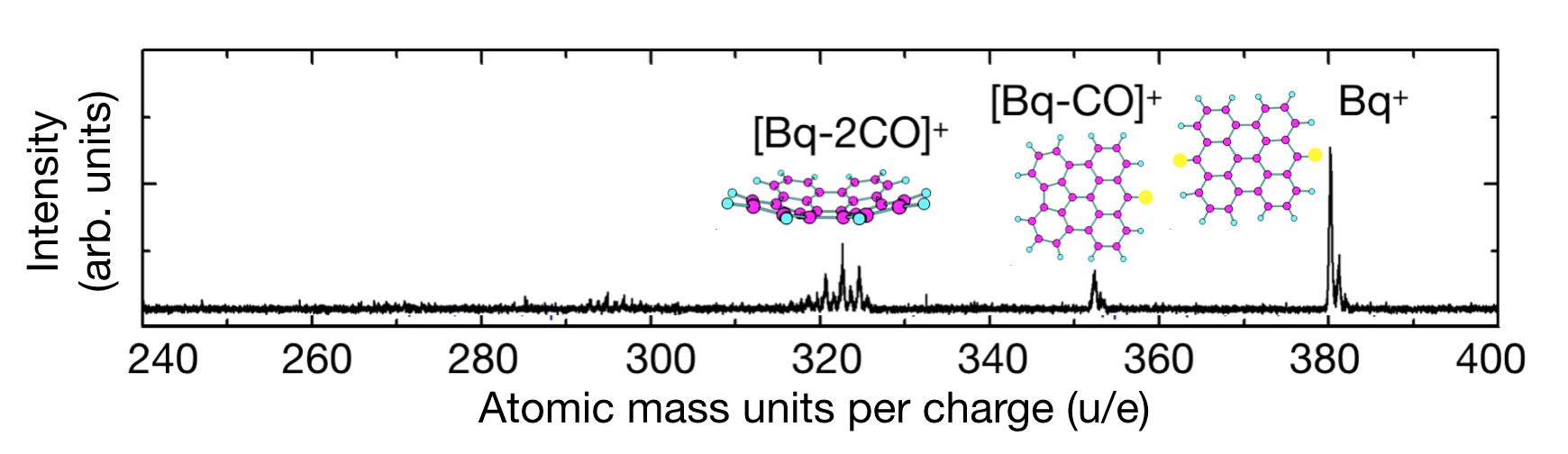}
\caption{Mass spectrum of bisanthenequinone cation (Bq$^+$) following irradiation by laser. The three major peaks respectively represent Bq$^+$, [Bq-CO]$^+$ and [Bq-2CO]$^+$ (from right to left). The optimized structures of the molecules are shown beside each peak. See \cite{chen2018planes} for details.}
\label{fig:bq}
\end{center}
\end{figure}

Figure~\ref{fig:pentagon} shows another example of pentagon formation in the N-substituted coronene. At 0.325\,ps, a HCN molecule is dissociated from the maternal molecule, which opens two hexagons and form two short carbon chains. At 0.335\,ps, one carbon chain closes up to form a pentagon. Then at 0.395\,ps another pentagon is formed due to the close up of the other carbon chain. These processes are similar to the formation of pentagons in Bq$^+$\citep{chen2018planes}. Recent work has also shown that the loss of a HCN molecule in nitrogen-containing PAHs offers a facile pathway towards pentagon formation \citep{de2017facile}.

%%% Figure 9 %%%
\begin{figure}
\begin{center}
\includegraphics[width=1.0\columnwidth]{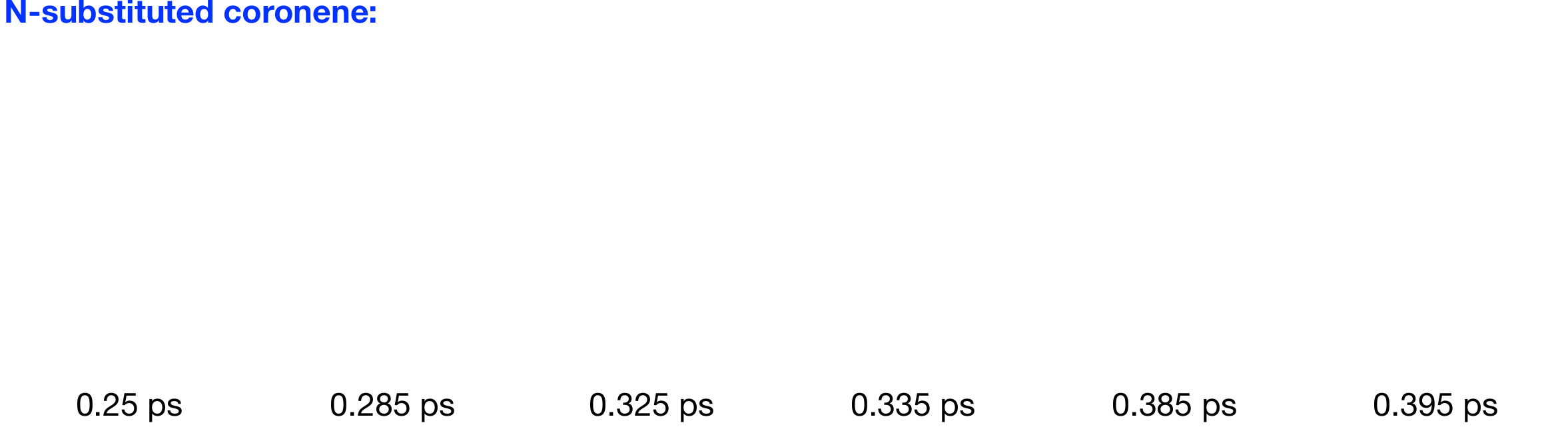}
\caption{Formation of pentagons in nitrogen-substituted coronene.}
\label{fig:pentagon}
\end{center}
\end{figure}

Apart from pentagons, we find the formation of heptagon in 1-methoxycoronene. Following the loss of two H$_2$ molecules, an unsaturated carbon (C=C) and a --CO functional group are formed (see Figure~\ref{fig:heptagon} and the last line in Figure~\ref{fig:fragmentation_4000K}). At 19.5\,ps, the bond linking the C=C and --CO units breaks. As shown above, the unsaturated carbon is very reactive, which always tends to form covalent bonds with a neighboring atom. At 19.53\,ps, the unsaturated carbon atom bends towards the neighboring carbon atom, and at 19.54\,ps, a heptagon is formed. 

%%% Figure 10 %%%
\begin{figure}
\begin{center}
\includegraphics[width=1.0\columnwidth]{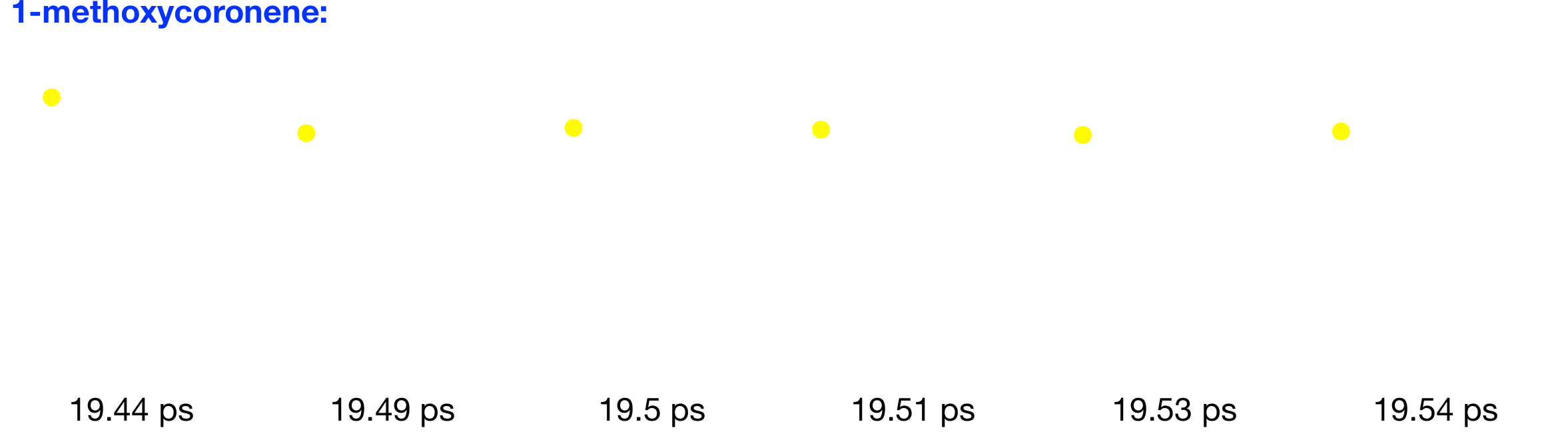}
\caption{Formation of heptagons in 1-methoxycoronene.}
\label{fig:heptagon}
\end{center}
\end{figure}

\section{Concluding Remarks}
To date, more than 200 molecules have been identified in the ISM. 
To understand the origin and evolution of astronomically-relevant
molecules, the time-of-flight (TOF) mass spectroscopy experiments in
combination with the static quantum chemical calculations are commonly
performed. However, such combination can hardly predict intermediate
or temporary products/states of the studied molecules, especially when
the molecules are vibrationally excited by photons, electrons or
stellar winds from nearby stars. 
In this work, 
using molecular dynamics simulations, various fragmentation and
isomerization features of coronene and its seven functionalized
derivatives have been investigated at high temperatures. 
It is found that most of these molecules do not dissociate at 3000$\K$, 
except hydrogenated coronene and oxygen-substituted coronene. 
These two molecules release hydrogen atoms and CO units at 3000\,K. 
Due to the loss of hydrogen, one of the peripheral ring
in hydrogenated coronene breaks, 
i.e., a six-ring attached with a --C$_2$H unit structure is formed. 
However, the structure of the oxygen-substituted coronene 
remains closed (i.e., no open rings) following the loss of CO. 
The other six molecules dissociate at 4000$\K$: 
(i) bare coronene releases a H$_2$ molecule 
without breaking any ring; 
(ii) 1-methylcoronene releases 
a CH$_2$ unit and forms a pentagon 
bonded with a --CH$_2$ unit at the end of the simulation; 
(iii) dehydrogenated coronene 
(i.e., coronene with a H atom loss) 
breaks completely to H atoms and carbon chains; 
(iv) nitrogen-substituted coronene releases a HCN unit 
and breaks two rings; 
(v) coronene with a H atom substituted by an oxygen atom 
dissociates to H, CO and carbon chains of various sizes; and 
(vi) 1-methoxycoronene releases two H$_2$ molecules 
and one CO unit, however, no ring breaks 
till the end of the simulation. 
The ring structures with carbon-chain 
attached and the formations of pentagons and heptagons 
are commonly observed in the simulations, 
especially in heteroatomic substituted molecules. 
The heteroatomic substitutions 
(e.g., nitrogen- and oxygen-substitutions) 
play a key role in the formation of pentagons or heptagons. 
We note that the vibrational temperatures of 3000 and 4000$\K$
examined in this work can not be easily achieved by small PAHs
through the absorption of single stellar photons
[see Figure~13 of \cite{draine2001infrared} and
Figure~4 of \cite{draine2007infrared}]. 
For coronene to attain such high vibrational temperatures,
it has to be excited by multiple photons or by energetic particles.
Therefore, the processes elaborated in this work mainly apply
to PAHs in UV-intense regions 
where they are excited by multiple photons 
or in hostile environments 
where they are collisionally excited by energetic particles.
Also, this is further complicated by 
the fact that coronene should be fully dehydrogenated in PDRs (see \citealt{andrews2016hydrogenation}) 
and even in the diffuse ISM (see \citealt{montillaud2013evolution}).

\section*{Acknowledgements}
We thank the referee for his/her vey helpful comments and suggestions.
This work is supported by the Swedish Research Council (Contract No. 2015-06501). The calculations were performed on resources provided by the Swedish National Infrastructure for Computing (SNIC) at the High Performance Computing Center North (HPC2N). AL is supported in part by NASA 80NSSC19K0572 and  NSF AST-1816411.

%\bibliographystyle{aa}
%\bibliography{../../savedrecs}

\end{document}